\documentclass[preprint,superscriptaddress,amssymb,pra,aps]{revtex4-2}
\usepackage{graphicx,xcolor,transparent,mathtools,amsmath,amsthm,soul,changes}
\usepackage{siunitx}
\begin{document}
\title{Coexistence of CHSH Nonlocality and KCBS Contextuality in a Single Quantum State
}

\author{Khai Nguyen}
\affiliation{Faculty of Physics, VNU University of Science, Vietnam National University, Hanoi, 120401, Viet Nam}
\affiliation{Institute for Quantum Technologies, Technology and Innovation Park, VNU Hanoi 120401, Vietnam}

\author{Duc M. Doan}
\affiliation{Faculty of Physics, VNU University of Science, Vietnam National University, Hanoi, 120401, Viet Nam}
\affiliation{Institute for Quantum Technologies, Technology and Innovation Park, VNU Hanoi 120401, Vietnam}
\author{Hung Q. Nguyen}
\email{hungngq@hus.edu.vn}
\affiliation{Faculty of Physics, VNU University of Science, Vietnam National University, Hanoi, 120401, Viet Nam}
\affiliation{Institute for Quantum Technologies, Technology and Innovation Park, VNU Hanoi 120401, Vietnam}
 
\date{\today}

\begin{abstract}

Contextuality and nonlocality are distinct manifestations at the foundation of quantum mechanics, yet their coexistence within a single quantum state remains subtle. In a hybrid CHSH--KCBS scenario involving the entanglment of a qubit and a qutrit,  the qutrit supports the KCBS contextuality test, and the CHSH nonlocality arises from correlations between the qubit and qutrit. Here, we derive the analytical closed-form expressions for both inequalities and also simulate this physics on a quantum circuit. We show that contextuality is governed solely by a population parameter $p_2$, associated with the occupation of the qutrit subsystem in the $|2\rangle$ level, which plays a distinguished role in the KCBS structure. In contrast, nonlocality depends irreducibly on coherence, involving both amplitudes and phases encoded in parameters $(X_i, Y_i)$. This separation of physical resources reveals  parameter regimes that optimize KCBS violation while suppress CHSH violation, and vice versa. As a result, the optimal regions do not overlap, and coexistence is restricted to a narrow intermediate regime in parameter space.
\end{abstract}

\maketitle

\section{Introduction}

Quantum theory exhibits forms of correlations that defy any classical 
description of physical reality. Among these, \emph{nonlocality} and 
\emph{contextuality} represent two fundamental manifestations of quantum 
nonclassicality, challenging locality and contextuality value assignments, 
respectively~\cite{Brunner2014,Budroni2022}. While both phenomena originate from the incompatibility between quantum 
observables and classical hidden-variable models, their relationship 
remains subtle. In particular, it is still unclear whether a single quantum 
system can simultaneously optimize both resources, or whether intrinsic 
constraints enforce a trade-off between them~\cite{Kurzynski2014,Araujo2024}. 

Quantum nonlocality traces back to the Einstein--Podolsky--Rosen (EPR) 
paradox~\cite{Einstein1935}, which questioned the completeness of quantum 
mechanics. Bell's theorem~\cite{Bell1964} subsequently translated this 
conceptual tension into experimentally testable inequalities that any local 
hidden-variable theory must satisfy. The Clauser--Horne--Shimony--Holt (CHSH) 
inequality~\cite{Clauser1969} provides the most widely used formulation: for 
dichotomic observables $\mathcal{A}_0,\mathcal{A}_1$ and 
$\mathcal{B}_0,\mathcal{B}_1$,
\begin{equation}
|\langle \mathcal{A}_1 \mathcal{B}_1 \rangle 
+ \langle \mathcal{A}_1 \mathcal{B}_0 \rangle 
+ \langle \mathcal{A}_0 \mathcal{B}_1 \rangle 
- \langle \mathcal{A}_0 \mathcal{B}_0 \rangle| 
\le 2. \label{CHSH_intro}
\end{equation}
Quantum mechanics violates this bound up to the Tsirelson limit 
$2\sqrt{2}$~\cite{cirel1980quantum}, a prediction confirmed in numerous 
experiments ranging from early tests~\cite{Aspect1982} to modern 
loophole-free demonstrations~\cite{Hensen2015,Shalm2015,Handsteiner2017}. 

In contrast, quantum contextuality does not rely on spatial separation. 
The Kochen--Specker theorem~\cite{KochenSpecker1967} proves that it is 
impossible to assign noncontextual definite values to quantum observables 
in a way consistent with all measurement contexts. While state-independent 
contextuality has been demonstrated through various constructions 
~\cite{mermin1990simple, yu2012state, cabello2015necessary, xu2015state}, 
a more operational and experimentally accessible approach is provided by 
noncontextuality inequalities. More generally, such inequalities can be 
formulated within the framework of $n$-cycle compatibility structures, 
which have been fully characterized and shown to be tight for arbitrary $n$ 
~\cite{araujo2013all, liang2011specker}. 

In particular, the case $n=5$ corresponds to the 
Klyachko--Can--Binicio\u{g}lu--Shumovsky (KCBS) inequality~\cite{Klyachko2008}, 
which captures state-dependent contextuality in the minimal qutrit scenario.
For an odd cycle of $n$ observables and $\theta_j = \frac{j(n-1)\pi}{n}$, define
\begin{equation}
|\psi_j\rangle = \frac{1}{\sqrt{1+\cos(\pi/n)}}
\Big[ \cos\theta_j,\ \sin\theta_j,\ \sqrt{\cos(\pi/n)} \Big]^T,
\end{equation}
and
\begin{equation}
B_j = (-1)^j \left( 2|\psi_j\rangle\langle\psi_j| - I \right),
\qquad \langle \psi_j | \psi_{j+1} \rangle = 0, \label{KCBS_Observable}
\end{equation}
so that adjacent observables commute. Noncontextual models impose
\begin{equation}
\Big\langle \sum_{j=0}^{n-2} B_j B_{j+1} - B_{n-1} B_0 \Big\rangle 
\le n - 2, \label{KCBS_ineq}
\end{equation}
which corresponds to the independence number $\alpha(C_n)$, while quantum 
theory reaches the Lov\'asz number $\vartheta(C_n)>\alpha(C_n)$ for odd 
cycles~\cite{cabello2014graph}. The minimal case $n=5$ already yields a 
qutrit contextuality violation~\cite{Klyachko2008}. 

Despite their common origin, nonlocality and contextuality were historically 
developed within distinct frameworks. A key conceptual advance was made by 
Cabello~\cite{Cabello2010}, who showed that nonlocal correlations can be 
understood as arising from local contextuality, a prediction later confirmed 
experimentally~\cite{Liu2016}. This connection has been further explored 
through monogamy relations, which suggest a trade-off between the two 
resources~\cite{Kurzynski2014}. However, more recent works indicate that 
such trade-offs are not universal, and that both resources may coexist in a 
single quantum state, albeit often in an unbalanced manner~\cite{Xue2023,
Kitajima2024}. 

From a broader perspective, contextuality has been increasingly recognized 
as a unifying resource underlying different forms of quantum correlations, 
bridging single-system contextuality and multipartite nonlocality~\cite{Budroni2022}. 
At the same time, quantum information platforms provide controlled settings 
in which both phenomena can be engineered and probed within circuit-based 
architectures. These developments naturally raise the question of whether a 
systematic and analytically tractable description of their interplay can be 
established.

In this work, we investigate a hybrid CHSH--KCBS scenario in a qubit--qutrit system, providing a unified analytical treatment of both inequalities. We derive closed-form expressions, determine the corresponding violation thresholds, and identify the parameter regimes governing their coexistence and competition. We further implement this scheme on quantum circuits to validate the analytical predictions. Our results reveal that the two violations are driven by distinct physical resources, leading to a constrained coexistence structure within the same quantum system.



\section{Locality and Noncontextuality Inequalities in a Hybrid Scenario}

\begin{figure}
    \centering
    \includegraphics[width=1\linewidth]{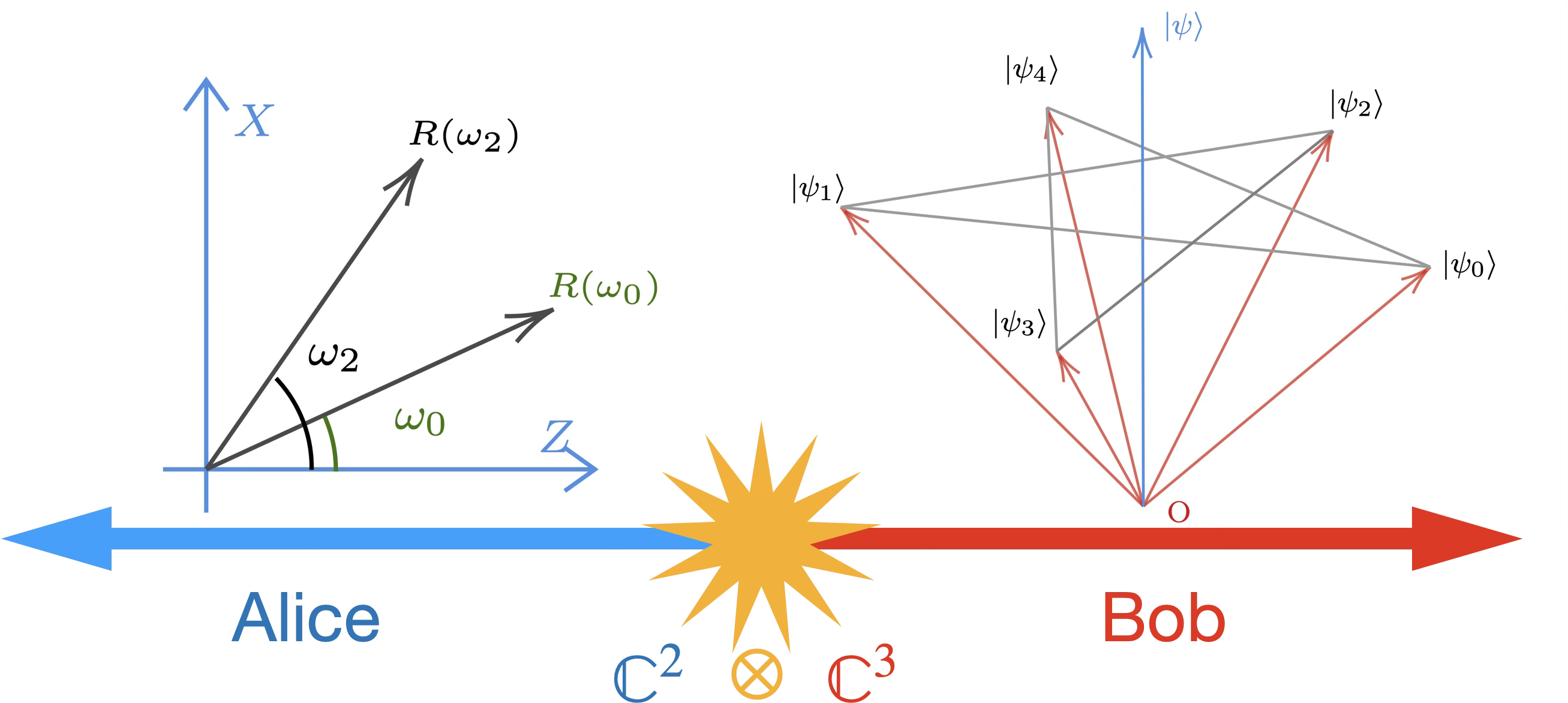}
    \caption{\textbf{Observing nonlocality and contextuality in $\mathbb{C}^2 \otimes \mathbb{C}^3$}. To observe nonlocality (CHSH), Alice performs measurements $R(\omega_0)$ and $R(\omega_2)$, while Bob measures observables $B_0$ and $B_m B_{m+1}$ and violate Eq.\eqref{CHSH_first}. To observe contextuality (KCBS), Alice performs no measurement, and Bob measures compatible observables $B_j$ and $B_{j+1}$ to evaluate the correlators $\langle B_j B_{j+1} \rangle$. He then obtains Eq.\eqref{KCBS_ineq}.}
    \label{AliceAndBob}
\end{figure}

The locality inequality and the noncontextuality inequality in a single system are defined in the Hilbert space $\mathbb{C}^2 \otimes \mathbb{C}^3$, where the most general pure state writes
\begin{equation}
    |\phi\rangle = \sum_{j=0}^{1} \sum_{k=0}^{2} c_{jk} |j\rangle |k\rangle,  \label{generalstate}
\end{equation}
with normalization condition $\sum_{j=0}^{1} \sum_{k=0}^{2} c_{jk}^* c_{jk} = 1.$

To construct the \textit{locality inequality}, and in analogy with the standard CHSH inequality, the following observables are defined, 
\begin{align*}
    \mathcal{A}_0 &= R(\omega_0) = \begin{pmatrix}
        \cos \omega_0 & \sin \omega_0\\ \sin \omega_0 & -\cos \omega_0
    \end{pmatrix} = Z \cos \omega_0 + X\sin \omega_0,\\
    \mathcal{A}_1 &= R(\omega_2) = \begin{pmatrix}
        \cos \omega_2 & \sin \omega_2\\ \sin \omega_2 & -\cos \omega_2
    \end{pmatrix} = Z \cos \omega_2 + X\sin \omega_2, \\
    \mathcal{B}_0 &= B_0 = \begin{pmatrix}
\dfrac{1-c}{1+c} & 0 &\dfrac{2\sqrt c}{1+c}\\
0&-1&0\\
\dfrac{2\sqrt c}{1+c}&0&\dfrac{c-1}{1+c}
\end{pmatrix}, \\
    \mathcal{B}_1 &= B_mB_{m+1} =  \begin{pmatrix} \frac{1-3c}{1+c} & 0 & \frac{4(-1)^m s_2 \sqrt{c}}{1+c} \\ 0 & 1 & 0 \\ \frac{4(-1)^m s_2 \sqrt{c}}{1+c} & 0 & \frac{3c-1}{1+c}\end{pmatrix}.
\end{align*}
With $c = \cos(\pi/n)$, $s_2 = \sin(\pi/2n)$, $m = \frac{n-1}{2}$, and $\omega_0,\omega_2$ parameterizing measurement directions in the $XZ$ plane of the Bloch sphere, as illustrated in Fig.\ref{AliceAndBob}.

The CHSH expression in Eq.~\eqref{CHSH_intro} is then given by
\begin{equation}
    |\langle R(\omega_2) \otimes B_{m}B_{m+1} \rangle + \langle R(\omega_2) \otimes B_0 \rangle + \langle R(\omega_0) \otimes B_{m}B_{m+1} \rangle - \langle R(\omega_0) \otimes B_0 \rangle|\overset{\text{LHV}}{\le} 2. \label{CHSH_first}
\end{equation}

As shown in Appendix ~\ref{calculate} and Eq.\eqref{CHSH_appendix}, the LHS of CHSH expression in Eq.\eqref{CHSH_first} can be written as
\begin{equation*}
\langle S_{\mathrm{CHSH}} \rangle = X_0 \cos\omega_0 + Y_0 \sin\omega_0 
+ X_2 \cos\omega_2 + Y_2 \sin\omega_2.
\end{equation*}
Here, the coefficients $\{X_0, Y_0, X_2, Y_2\}$ are determined by the state amplitudes and phases
\begin{equation}
\begin{aligned}
X_0 &= \frac{-2c}{1+c} \left( |c_{00}|^2 - |c_{10}|^2 - |c_{02}|^2 + |c_{12}|^2 \right) + 2 \left( |c_{01}|^2 - |c_{11}|^2 \right) \\ &  \qquad \qquad \qquad \qquad \qquad \qquad \qquad \qquad \qquad   + \frac{2\sqrt{c}}{1+c} \text{Re}(c_{00}^* c_{02} - c_{10}^* c_{12}) \left( 4(-1)^m s_2 - 2 \right), \\
Y_0 &= \frac{-4c}{1+c} \text{Re}(c_{10}^* c_{00} - c_{12}^* c_{02}) + 4\text{Re}(c_{11}^* c_{01}) + \frac{2\sqrt{c}}{1+c} \text{Re}\left( c_{12}^* c_{00} + c_{02}^* c_{10} \right) \left( 4(-1)^m s_2 - 2 \right), 
\end{aligned} \label{X0Y0_0}
\end{equation}
and
\begin{equation}
\begin{aligned}
X_2 &= \frac{2-4c}{1+c} \left( |c_{00}|^2 - |c_{10}|^2 - |c_{02}|^2 + |c_{12}|^2 \right) + \frac{2\sqrt{c}}{1+c} \text{Re}(c_{00}^* c_{02} - c_{10}^* c_{12}) \left( 4(-1)^m s_2 + 2 \right), \\
Y_2 &= 2 \left( \frac{2-4c}{1+c} \right) \text{Re}(c_{10}^* c_{00} - c_{12}^* c_{02}) + \frac{2\sqrt{c}}{1+c} \text{Re}\left( c_{12}^* c_{00} + c_{02}^* c_{10} \right) \left( 4(-1)^m s_2 + 2 \right).
\end{aligned} \label{X2Y2_0}
\end{equation}
With optimal settings
$\omega_i = \arctan\!\left(\frac{Y_i}{X_i}\right)$, for $i=0,2$, the CHSH expression reduces to a sum of two independent contributions, each associated with an effective Bloch vector in the $Z$–$X$ plane, with optimal measurements aligned along these directions.

A violation occurs whenever
\begin{equation}
S_{\rm{CHSH}}^{\rm{opt}}=\sqrt{X_0^2 + Y_0^2} + \sqrt{X_2^2 + Y_2^2} > 2. \label{CHSH_opti}
\end{equation}
This condition does not reduce to a simple threshold on a single population. Nonlocality emerges from a nontrivial combination of populations and coherences encoded in $(X_i, Y_i)$, reflecting its intrinsically bipartite and interference-driven origin.

We now construct the \textit{noncontextuality inequality}. Here, Alice performs no measurement, while Bob selects a measurement context indexed by $i \in \{0,\dots,n-1\}$ and measures compatible observables $B_i$ and $B_{i+1}$, as introduced in Eq.\eqref{KCBS_Observable}.
The corresponding noncontextuality inequality reads
\begin{equation}
\left\langle I_2 \otimes \left( \sum_{j=0}^{n-2} B_j B_{j+1} - B_{n-1} B_0 \right) \right\rangle \overset{\text{NCHV}}{\le} n-2.
\label{KCBS_2}
\end{equation}

A violation of the noncontextuality inequality requires that the left-hand side of Eq.~\eqref{KCBS_2}, namely the expectation value $S_{\mathrm{KCBS}}$, satisfies
\begin{equation}
    S_{\rm{KCBS}} = n\frac{4c-2}{1+c}\sum_{j=0}^1 |c_{j2}|^2 + n\frac{1-c}{1+c} > n-2. \label{KCBS_SUM}
\end{equation}
As usual $c = \cos(\pi/n)$, this condition is equivalent to
\begin{equation}
\sum_{j=0}^1 |c_{j2}|^2 = p_2  
> \frac{nc - 1 - c}{(2c - 1)n},
\label{thresholdKCBS}    
\end{equation}
see Appendix \ref{calculate} for detail derivation. 
Our result implies that any state with a sufficiently large population in the $|2\rangle$ level of Bob's subsystem exhibits contextuality. For instance, for $n = 5, 7, 9$, and 11, one finds 
$p_2 > 0.724, \, 0.785,\, 0.824$, and 0.850, respectively, 
showing that the threshold $p_2$ increases monotonically with $n$ and approaches unity.


The two inequalities Eq.\eqref{CHSH_opti} and Eq. \eqref{thresholdKCBS} impose distinct physical requirements: CHSH violations 
require entanglement and inter-subsystem coherence, while KCBS violations 
necessitate nonzero population in Bob’s $|2\rangle$ level. Starting from a general state in $\mathbb{C}^2 \otimes \mathbb{C}^3$, symmetry 
and phase redundancies allow a reduction to a minimal parametrization in terms 
of two control variables, suitable for both circuit implementation and 
visualizing the violation landscape.

As a minimal CHSH--KCBS state, the state
\begin{equation}
|\psi_1\rangle = \sin\frac{\theta}{2}\,|00\rangle 
+ \cos\frac{\theta}{2}\, e^{i\phi}|12\rangle
\label{stateI}
\end{equation} captures the minimal structure required for the coexistence of the 
two resources: coherence between subsystems, responsible for nonlocality, and 
population in the $|2\rangle$ level, required for contextuality. Up to local 
basis choices and global phases, it provides a canonical representative of the 
hybrid CHSH--KCBS scenario.

This minimal ansatz is sufficient to reveal the interplay and trade-off between 
the two types of violations while remaining directly compatible with a simple 
quantum circuit implementation.

\section{Simulation on Quantum Circuits}

\begin{figure}
\centering
\includegraphics[width=0.9\linewidth]{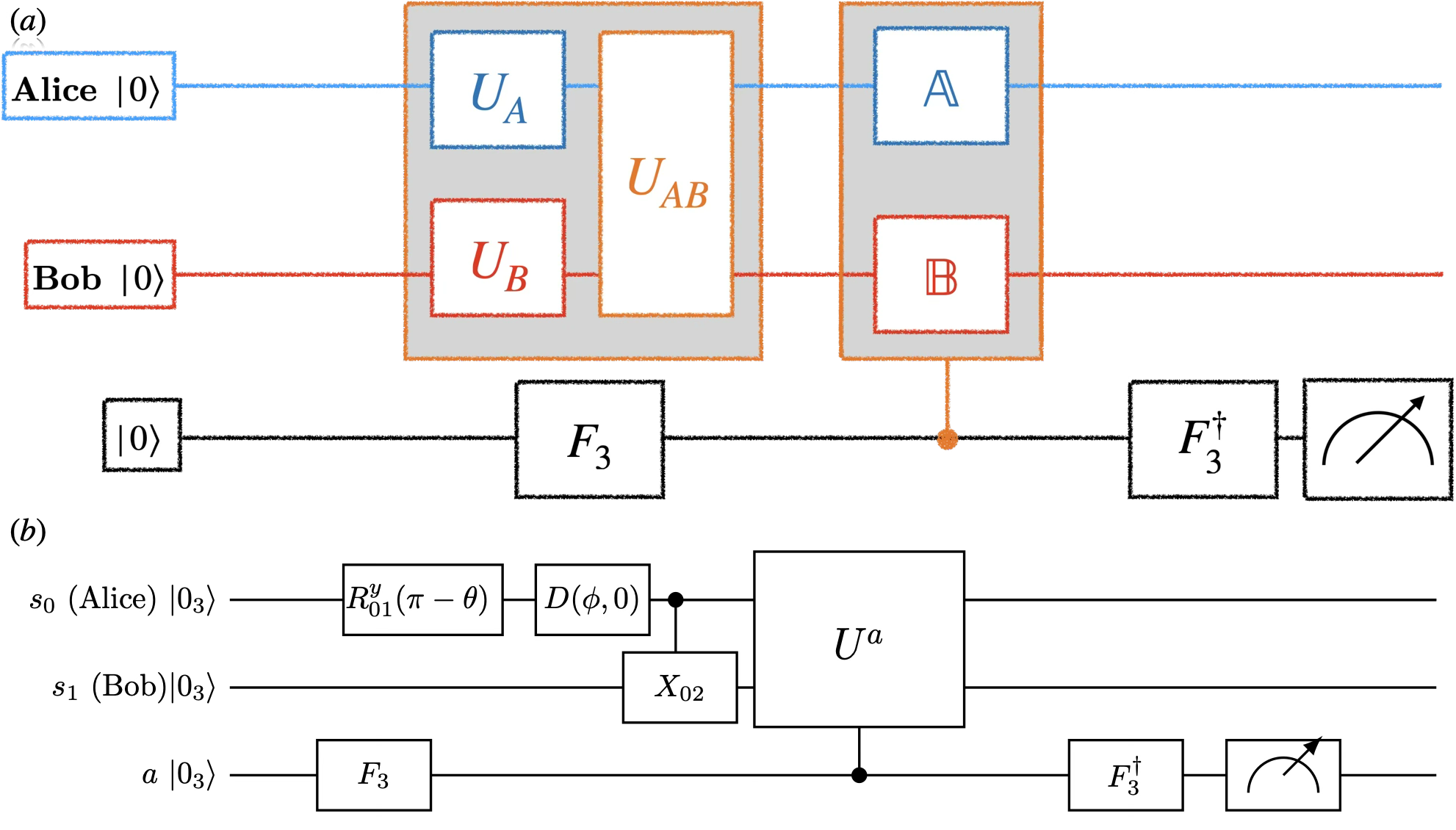}
    \caption{
    Quantum circuit realization of the hybrid CHSH--KCBS protocol. (a) General architecture: the first stage prepares the quantum state, while the second stage implements the measurement via a qutrit Fourier test. The controlled-$U^a$ operation encodes the joint observable $U = A \otimes B$, where $A$ and $B$ denote Alice's and Bob's measurement operators. (b) Explicit implementation for the state in Eq.~\eqref{stateI}. All gate conventions (e.g., $F_3$, $R_{ij}(\theta)$, $D(\alpha,\beta)$, and controlled-$U^a$) follow Sec.~\ref{Gates}.}
    \label{QCircuit}
\end{figure}

Motivated by the Hadamard test circuit introduced in~\cite{NielsenChuang}, we propose a Fourier test circuit for qutrit systems. As shown explicitly in Appendix ~\ref{QC_Fourier_Proof} and Eq.\eqref{U_expectation}, this circuit provides a method to evaluate the expectation value of a unitary and Hermitian operator \( U \) in a qutrit system. 

Specifically, by measuring the ancilla qutrit, the expectation value \( \langle U \rangle \) can be obtained as
\begin{equation}
\langle U\rangle
=
\frac{9P(0) - 5}{4}
=
\frac{2 - 9P(1)}{2} =
\frac{2 - 9P(2)}{2}
=
\frac{9\big(P(0) - P(1) - P(2)\big) - 1}{8} 
\label{eq:U_expectation}
\end{equation}
where \( \{P(0), P(1), P(2)\} \) denote the measurement probabilities of the ancilla qutrit. 
Thus, the expectation value is operationally obtained via the general Fourier test circuit shown in Fig\ref{QCircuit}.a.
A unified circuit framework for implementing the hybrid CHSH--KCBS protocol is illustrated in Fig.~\ref{QCircuit}. Here, figure~\ref{QCircuit}(a) depicts the general architecture, applicable to arbitrary input states, while Figs.~\ref{QCircuit}(b,c) provide explicit realizations for the two states considered in this work. The protocol is formulated in an effective two-qutrit Hilbert space. Alice’s subsystem is restricted to the qubit subspace $\{|0\rangle, |1\rangle\}$, whereas Bob’s subsystem fully occupies the qutrit space $\{|0\rangle, |1\rangle, |2\rangle\}$.

The construction follows a unified two-step procedure, schematically illustrated in Fig.~\ref{QCircuit}(a):

\vspace{0.2cm}
\noindent \textbf{Step 1: State preparation.}
The first stage [Fig.~\ref{QCircuit}(a)] prepares the target state via local unitaries $U_A$, $U_B$ and an entangling operation $U_{AB}$, which together generate the bipartite state used in the protocol.

An explicit implementation for the state in Eq.~\eqref{stateI} is shown in Fig.~\ref{QCircuit}(b). It consists of three gates: a rotation $R_{01}^y(\pi - \theta)$, a phase gate $D(\phi,0)$, and a controlled-$X_{02}$ operation, which together prepare the desired superposition between $|00\rangle$ and $|12\rangle$. Further details are provided in Appendix~\ref{Preparingstate1}.

\vspace{0.2cm}
\noindent \textbf{Step 2: Measurement stage.}
As summarized in Fig.~\ref{QCircuit}(a), the measurement layer determines the effective observable $U = \mathbb{A} \otimes \mathbb{B} $ implemented in the Fourier test circuit.

\begin{itemize}
    \item \textbf{CHSH scenario:} 
Alice and Bob independently select their local observables $\mathbb{A} = \{ R(\omega_0),R(\omega_2) \}$ and $\mathbb{B}= \{ B_0,B_mB_{m+1} \}$, defining a bipartite operator $U =\mathbb{A} \otimes \mathbb{B}$.
    \item \textbf{KCBS scenario:} 
Alice performs no measurement ($\mathbb{A} = I$), while Bob selects compatible observable pairs  $\mathbb{B} = B_j B_{j+1}$, yielding $U = I \otimes (B_j B_{j+1})$.
\end{itemize}

Finally, following the general structure in Fig.~\ref{QCircuit}(a), the controlled unitary $U^a$ in the Fourier test circuit is configured according to the chosen measurement setting, allowing direct estimation of the expectation value $\langle U \rangle$.

\section{Simulation and Analyzing of a state exhibiting both nonlocality and contextuality}

\begin{figure}
    \centering
    \includegraphics[width=1\linewidth]{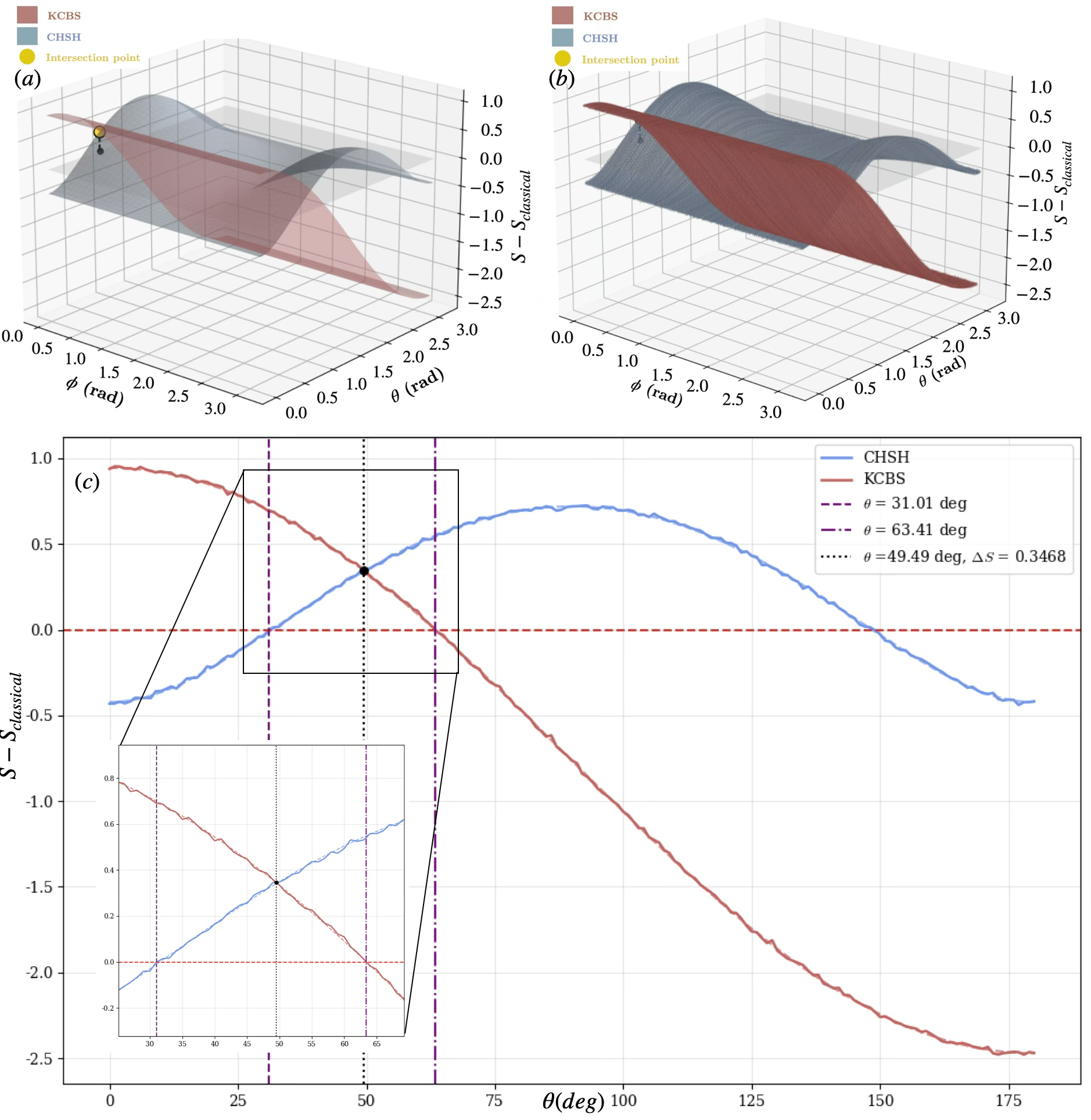}
    \caption{A state exhibiting both nonlocality and contextuality. 
    (a) Analytical violation landscape for the hybrid CHSH-KCBS scenario. 
    (b) Circuit-based simulation, showing excellent agreement with the analytical results. 
    (c) One-dimensional slice at $\phi = 0$ (or $\phi = k\pi$), corresponding to the maximal CHSH violation. 
    The intersection point identifies the optimal coexistence between nonlocality and contextuality.}
    \label{fig:state1}
\end{figure}

The chosen state in Eq.~\ref{stateI} spans an effective two-dimensional subspace of $\mathbb{C}^2 \otimes \mathbb{C}^3$. From Eq.~(\ref{CHSH_appendix}), the relevant coefficients are
\[
\begin{aligned}
X_0 &= -\frac{2c}{1+c}, \\
Y_0 &= \frac{\sqrt{c}\,\sin\theta\cos\phi}{1+c}\left(4(-1)^m s_2 - 2\right), \\
X_2 &= \frac{2-4c}{1+c}, \\
Y_2 &= \frac{\sqrt{c}\,\sin\theta\cos\phi}{1+c}\left(4(-1)^m s_2 + 2\right).
\end{aligned}
\]

The optimized CHSH value and the KCBS value in Eq.\eqref{CHSH_opti} and  Eq.\eqref{KCBS_SUM} are then
\[
S_{\rm{CHSH}}^{\mathrm{opt}} = \sqrt{X_0^2 + Y_0^2} + \sqrt{X_2^2 + Y_2^2},
\qquad
S_{\rm{KCBS}} = \frac{n}{1+c}\left[(4c-2)\cos^2\left(\frac{\theta}{2}\right) + 1 - c\right].
\]

The corresponding violation margins are
\begin{equation}
\begin{aligned}
S_{\rm{CHSH}}^{\mathrm{opt}} - 2
&=
\frac{1}{1+c}
\left[
\sqrt{4c^2 + c\sin^2\theta\cos^2\phi\left(4(-1)^m s_2 - 2\right)^2}
\right. \\
&\hspace{2.5em}\left.
+
\sqrt{(2-4c)^2 + c\sin^2\theta\cos^2\phi\left(4(-1)^m s_2 + 2\right)^2}
\right] - 2,
\\[0.5em]
S_{\rm{KCBS}} - (n-2)
&=
\frac{n}{1+c}\left[(4c-2)\cos^2\left(\frac{\theta}{2}\right) - 2c\right] + 2.
\end{aligned}
\label{Sum_for_state_I}
\end{equation}

As seen in Figs.~\ref{fig:state1}(a,b), the analytical predictions are in excellent agreement with the circuit-based simulations, accurately reproducing both the violation landscape and the coexistence point. At the maximal intersection, the analytical solution gives $\phi = 0^\circ$ and $\theta \approx 49.65^\circ$, with a violation of approximately $0.34$, while the simulation yields $\theta \approx 50.8^\circ$, $\phi = 0^\circ$, and a violation of approximately $0.38$. This agreement confirms that the closed-form expressions capture the essential physics governing the joint violation.

The two violations are controlled by distinct physical resources. CHSH is driven by phase-sensitive coherence through the interference term $\sim \sin\theta\cos\phi$ and is therefore maximized at $\phi = k\pi$ and $\theta = \pi/2$, where the superposition between basis components is strongest. KCBS, by contrast, is purely geometric: it depends only on the population imbalance $\sim \cos^2(\theta/2)$ and reaches its maximum at $\theta = 0$, independent of the relative phase. Hence, the two inequalities favor different regions of the state space, so optimizing one necessarily moves the state away from the optimal regime of the other.
\begin{figure}
    \centering
    \includegraphics[width=1\linewidth]{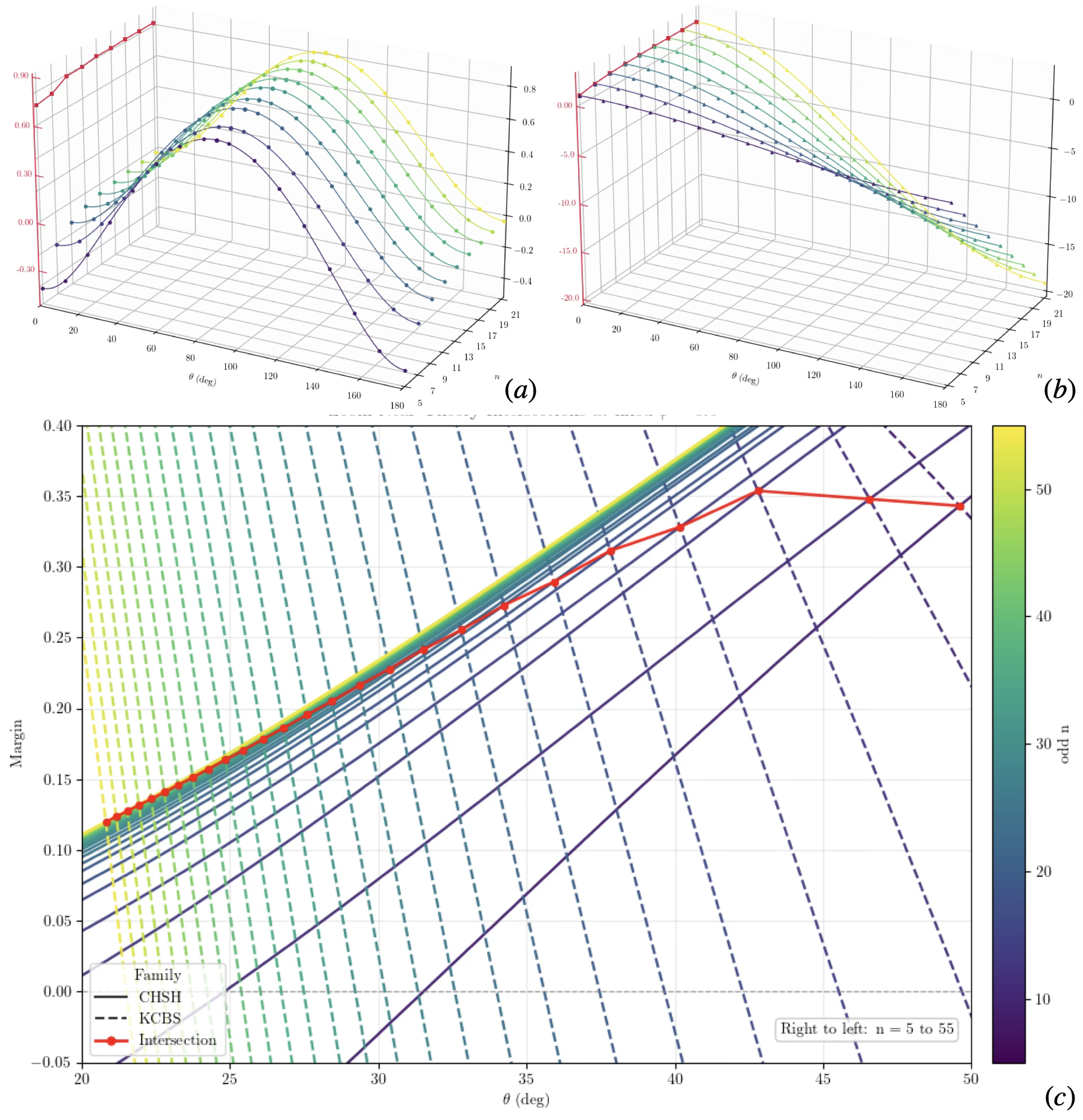}
    \caption{Scaling of the optimal coexistence point with the cycle size $n$ for State I.
    (a) CHSH expectation value as a function of $\theta$ for representative values of $n$ (with $\phi=0$).
    (b) Corresponding KCBS violation for the same set of $n$.
    (c) Intersection of the CHSH and KCBS violation margins $S - S_{\mathrm{classical}}$ for odd $n \in [5,55]$, defining the optimal coexistence point.
    The optimal angle $\theta_{\mathrm{opt}}$ decreases with $n$ (approximately $\sim n^{-1/2}$), while the maximal violation at the intersection diminishes, indicating a progressive weakening of the coexistence region.}
    \label{simulandanal_state1_intersection}
\end{figure}

As $n$ increases, the geometric enhancement strengthens, shifting the coexistence optimum toward smaller $\theta$ and away from the CHSH-favored regime near $\pi/2$. From Eq.~\ref{Sum_for_state_I} and Fig.~\ref{fig:state1}, both margins are maximized at $\phi = k\pi$; hence we set $\phi=0$ and optimize over $\theta$. Figures~\ref{simulandanal_state1_intersection}(a,b) and Table~\ref{data_state_1} show that the simultaneous violation region shrinks with $n$, while the optimal angle moves toward $\theta \to 0$. In the large-$n$ limit, the compromise scales as
\[
\theta^2 = \frac{8}{n+4} \sim \frac{1}{n},
\qquad
S_{\mathrm{quantum}} - S_{\mathrm{classical}} \sim \frac{8}{n+4} \sim \frac{1}{n} > 0.
\]
This behavior reflects the different physical resources involved: CHSH is coherence-driven, whereas KCBS is enhanced by population concentration on Bob’s side. Accordingly, CHSH is favored by highly coherent superpositions such as
\[
|\psi\rangle = \frac{1}{\sqrt{2}}\left(|00\rangle + e^{ik\pi}|12\rangle\right),
\]
while KCBS prefers states biased toward $|2\rangle$. Thus, although the coexistence region contracts with $n$, it remains nonempty for every finite $n$ (see Appendix \ref{large_n_state_I}).

Motivated by the optimal large-$n$ scaling $\theta^2 \sim \frac{8}{n+4}$, we construct a family of states whose population imbalance reproduces this behavior. We observe that within the measurement scenario defined above, for every odd integer $n \ge 5$ there exists a bipartite qutrit state that simultaneously violates the CHSH inequality and the $n$-cycle KCBS inequality.

A concrete realization is given by
\begin{equation}
|\psi_n\rangle = \sqrt{\frac{2}{n+4}}\,|00\rangle 
+ \sqrt{\frac{n+2}{n+4}}\, e^{ik\pi}|12\rangle , 
\label{Theorem1_eq}
\end{equation}
$k \in \mathbb{Z}$. With this choice, the violations admit the asymptotic forms
\begin{align}
S_{\mathrm{KCBS}} - (n-2) = \frac{8}{n+4} \sim \frac{1}{n} > 0,\\
S_{\mathrm{CHSH}} - 2 = \frac{8(n+2)}{(n+4)^2} \sim \frac{1}{n} > 0 ,
\end{align}
which are strictly positive for all odd $n \ge 5$. Detailed derivations are provided in Appendix~\ref{explicitTheorem1}.
The above expressions are obtained at leading order in the large-$n$ expansion. 
Subleading corrections from higher-order trigonometric expansions may slightly modify numerical prefactors, 
but do not affect the scaling behavior or the coexistence of violations.

\begin{table}
\caption{Optimal coexistence point as a function of the cycle size $n$. 
The table lists the optimal angle $\theta_{\mathrm{opt}}$ (in degrees) 
and the corresponding overlap.}
\label{data_state_1}
\begin{ruledtabular}
\setlength{\tabcolsep}{4pt}
\begin{tabular}{
c c c
c c c
c c c
}
\footnotesize
$n$ & $\theta_{\mathrm{opt}}$ & overlap
& $n$ & $\theta_{\mathrm{opt}}$ & overlap
& $n$ & $\theta_{\mathrm{opt}}$ & overlap \\
\hline
5  & 49.605 & 0.343069 & 23 & 30.381 & 0.227717 & 41 & 23.747 & 0.151422 \\
7  & 46.568 & 0.347839 & 25 & 29.355 & 0.216111 & 43 & 23.252 & 0.145882 \\
9  & 42.804 & 0.353697 & 27 & 28.432 & 0.205008 & 45 & 22.785 & 0.140814 \\
11 & 40.174 & 0.328131 & 29 & 27.588 & 0.195383 & 47 & 22.346 & 0.136011 \\
13 & 37.825 & 0.311358 & 31 & 26.818 & 0.186256 & 49 & 21.932 & 0.131586 \\
15 & 35.922 & 0.289453 & 33 & 26.107 & 0.178192 & 51 & 21.54  & 0.127384 \\
17 & 34.24  & 0.272828 & 35 & 25.452 & 0.170568 & 53 & 21.168 & 0.123487 \\
19 & 32.802 & 0.255515 & 37 & 24.843 & 0.163735 & 55 & 20.815 & 0.11978  \\
21 & 31.515 & 0.241466 & 39 & 24.277 & 0.157276 &     &        &          \\
\end{tabular}
\end{ruledtabular}
\end{table}

Our results reveal a genuine trade-off between contextuality and nonlocality: parameter regimes that enhance the KCBS violation suppress the CHSH violation, and vice versa. This is not a numerical artifact, but a structural competition between two distinct nonclassical resources within the same operational setting. The key question is whether this trade-off is intrinsic, in the sense that it is enforced by the structure of quantum correlations themselves, or whether it is contingent on the specific observables, state parametrization, and measurement configuration adopted here. Importantly, this competition is not universal. It arises within a state-dependent contextuality scenario; by contrast, state-independent contextuality would exhibit uniform violations across all states. The observed trade-off therefore reflects the present construction rather than contextuality as such.

\section{Coexistence of Contextuality and Nonlocality}

\begin{table}
\centering
\begin{tabular}{c|c|c}
\hline
\textbf{Quantity} & \textbf{Definition} & \textbf{Role} \\
\hline
$Q_0$ & $|c_{00}|^2 - |c_{10}|^2 - |c_{02}|^2 + |c_{12}|^2$ 
& Population \\
$Q_1$ & $|c_{01}|^2 - |c_{11}|^2$ 
& Population \\
\hline
$R_1$ & $\mathrm{Re}(c_{00}^* c_{02} - c_{10}^* c_{12})$ 
& Coherence \\
$R_2$ & $\mathrm{Re}(c_{10}^* c_{00} - c_{12}^* c_{02})$ 
& Coherence \\
$R_3$ & $\mathrm{Re}(c_{12}^* c_{00} + c_{02}^* c_{10})$ 
& Coherence \\
$R_4$ & $\mathrm{Re}(c_{11}^* c_{01})$ 
& Coherence \\
\hline
$S_{\pm}$ & $4(-1)^m s_2 \pm 2$ 
& Geometry \\
$c$ & $\cos(\pi/n)$ 
& Geometry \\
$s_2$ & $\sin(\pi/2n)$ 
& Geometry \\
\hline
\end{tabular}
\caption{Decomposition into population ($Q$), coherence ($R$), and geometric contributions.}
\label{Component}
\end{table}

We decompose the state contributions into two distinct sectors, \emph{population (diagonal)} and \emph{coherence (off-diagonal)}. The CHSH expression becomes a sum of two Euclidean norms:
\begin{equation*}
S_{\rm{CHSH}}^{opt} = \sqrt{X_0^2 + Y_0^2} + \sqrt{X_2^2 + Y_2^2},
\end{equation*}
where
\begin{align*}
X_0 &= \frac{-2c}{1+c} Q_0 + 2 Q_1 + \frac{2\sqrt{c}}{1+c} R_1 S_-, \\
Y_0 &= \frac{-4c}{1+c} R_2 + 4 R_4 + \frac{2\sqrt{c}}{1+c} R_3 S_-, \\
X_2 &= \frac{2-4c}{1+c} Q_0 + \frac{2\sqrt{c}}{1+c} R_1 S_+, \\
Y_2 &= \frac{4-8c}{1+c} R_2 + \frac{2\sqrt{c}}{1+c} R_3 S_+.
\end{align*}
The CHSH value is a quadratic form mixing population ($Q$) and coherence ($R$) contributions, where interference enters through $R_i$ and is geometrically modulated by $S_{\pm}$. In contrast, the KCBS expression depends purely on populations:
\begin{equation*}
S_{\rm{KCBS}} = \frac{n}{1+c} \left[ (4c-2)\sum_{j=0}^{1} |c_{j2}|^2 + (1-c) \right].
\end{equation*}

The origin of the trade-off becomes transparent at the level of the density matrix. The KCBS expression depends only on diagonal populations, specifically the weight in the subspace with Bob in \(|2\rangle\), and is therefore driven entirely by population redistribution. By contrast, the CHSH expression combines population imbalances \((Q_0,Q_1)\) with coherence terms \((R_i)\), where, for instance,
\begin{equation*}
Q_1 = |c_{01}|^2 - |c_{11}|^2, \qquad 
Q_0 = |c_{00}|^2 - |c_{10}|^2 - |c_{02}|^2 + |c_{12}|^2.
\end{equation*}
These quantities are not simple populations but imbalances, which can be strongly suppressed when weight is concentrated in a restricted sector.

These two requirements are structurally misaligned. Enhancing KCBS concentrates weight into the $| 2 \rangle $ sector, which directly reduces the contributions entering \(Q_0\) and suppresses the imbalance \(Q_1\), while also limiting the amplitude spread needed to sustain the coherences \(R_i\). Conversely, maintaining strong CHSH interference typically distributes the state more broadly across basis components, which weakens the population concentration that favors KCBS.


\section{Discussion}

We have investigated the interplay between CHSH nonlocality and KCBS contextuality within a unified qutrit-based architecture. By introducing a minimal state ansatz and deriving closed-form expressions for both inequalities, we obtain a transparent and analytically tractable description of how these two forms of nonclassicality coexist and compete at both the state and circuit levels.

A central result of this work is that the observed trade-off between nonlocality and contextuality is \emph{structural}, in the sense that it originates from how the two inequalities access complementary sectors of the density matrix. The KCBS expression depends exclusively on diagonal populations, in particular the weight in the subspace where Bob occupies $|2\rangle$, and is therefore governed by population redistribution. In contrast, the CHSH expression depends on both population imbalance and phase-sensitive coherences, requiring interference between multiple components. As a result, the two resources cannot be simultaneously optimized within the present setting.

Despite this intrinsic competition, simultaneous violation of both inequalities remains possible for all finite cycle sizes $n$. However, the coexistence region becomes progressively compressed as $n$ increases. In particular, the optimal regime shifts toward small angles, with $\theta_{\mathrm{opt}} \sim \mathcal{O}(1/\sqrt{n})$, indicating that the coherence required for CHSH violation is confined to an increasingly narrow region of parameter space, while contextuality is geometrically amplified by the $n$-cycle structure. This scaling behavior provides a clear physical picture of coexistence as a balance between geometric amplification and coherence suppression.

Importantly, this trade-off is not universal. It arises from the specific combination of state-dependent contextuality, the chosen compatibility structure, and the restricted class of states considered here. In more general scenarios, such as state-independent contextuality constructions, alternative measurement configurations, or higher-dimensional embeddings—the relationship between nonlocality and contextuality may differ substantially. The present results therefore indicate that the interplay between these resources is strongly framework-dependent rather than dictated by a general constraint of quantum theory.

From an implementation perspective, the proposed qutrit Fourier-test circuit provides a concrete and resource-efficient method for evaluating both inequalities within a single experimental platform. The ability to access CHSH and KCBS correlations by modifying only the measurement stage makes this approach particularly suitable for near-term quantum platforms, such as NISQ devices, where flexible and low-overhead characterization of nonclassical correlations is essential.

An open question raised by our analysis is whether the observed trade-off reflects a deeper constraint on the simultaneous optimization of distinct nonclassical resources, or whether it can be mitigated by more general state constructions or alternative compatibility structures. Addressing this question may contribute to a more unified understanding of quantum correlations and their role as fundamental operational resources in quantum information processing.

\bibliographystyle{unsrt}
\bibliography{references}

\appendix
\numberwithin{equation}{section}

\section{Derivations}
\subsection{General state for both Locality and Non-Contextuality inequality} \label{calculate}

The general state in $\mathbb{C}^2 \otimes \mathbb{C}^3$ is: \begin{equation*}
    |\phi\rangle = \sum_{j=0}^{1} \sum_{k=0}^{2} c_{jk} |j\rangle |k\rangle, \quad \sum_{j=0}^{1} \sum_{k=0}^{2} c_{jk}^* c_{jk} = 1. 
\end{equation*}
And their dual state: 
\begin{equation*}
    \langle\phi| = \sum_{j=0}^{1} \sum_{k=0}^{2} c_{jk}^* \langle k| \langle j|
\end{equation*}
Let $A \in \mathcal{H}_2$ and $B \in \mathcal{H}_3$ be Hermitian operators, where
\[
\mathcal{H}_n := \{ X \in M_n(\mathbb{C}) \mid X^\dagger = X \}.
\]
The expectation value of $A\otimes B$ on this state is: 

\[ S =  \langle A \otimes B \rangle =  
\left[ \sum_{e=0}^{1} \sum_{f=0}^{2} c_{ef}^* \langle f| \langle e| \right] A \otimes B \left[ \sum_{j=0}^{1} \sum_{k=0}^{2} c_{jk} |j\rangle |k\rangle \right]
\]\[
S = \sum_{e=0}^{1} \sum_{f=0}^{2} \sum_{j=0}^{1} \sum_{k=0}^{2} c_{ef}^* c_{jk} \langle e|A|j\rangle \langle f|B|k\rangle\]
In this scenario: $A = R(\omega)$
\[A = R(\omega) = \begin{pmatrix} \cos \omega & \sin \omega \\ \sin \omega & -\cos \omega \end{pmatrix} \Rightarrow  
\begin{aligned}
&\langle 0|A|0\rangle = \cos \omega \ ; \ \langle 1|A|1\rangle = -\cos \omega \\
&\langle 1|A|0\rangle = \langle 0|A|1\rangle = \sin \omega 
\end{aligned}\]
In short: 
\begin{equation}
S = \sum_{f=0}^{2} \sum_{k=0}^{2} \underbrace{\left[ \cos \omega (c_{0f}^* c_{0k} - c_{1f}^* c_{1k}) + \sin \omega (c_{0f}^* c_{1k} + c_{1f}^* c_{0k}) \right]}_{D_{fk}} \langle f|B|k\rangle  
\end{equation}
\textbf{Now, recall matrix $B_mB_{m+1}$ in \ref{BmBm+1}: } \\
\[B_m B_{m+1} = \begin{pmatrix} 
\frac{1-3c}{1+c} & 0 & \frac{4(-1)^m s_2 \sqrt{c}}{1+c} \\ 
0 & 1 & 0 \\ 
\frac{4(-1)^m s_2 \sqrt{c}}{1+c} & 0 & \frac{3c-1}{1+c} 
\end{pmatrix} \Rightarrow
\begin{aligned}
&\langle 0|B|0\rangle = -\langle 2|B|2\rangle = \frac{1-3c}{1+c} \\
&\langle 1|B|1\rangle = 1 \\
&\langle 2|B|0\rangle = \langle 0|B|2\rangle = \frac{4(-1)^m s_2 \sqrt{c}}{1+c}
\end{aligned}
\]
Then
\[
S_{B_mB_{m+1}} = (D_{00} - D_{22}) \frac{1-3c}{1+c} + D_{11} + (D_{02} + D_{20}) \frac{4(-1)^m s_2 \sqrt{c}}{1+c}\]

\textbf{Now, recall matrix $B_0$ in \ref{B0}: }

\[B_0 = \begin{pmatrix} 
\frac{1-c}{1+c} & 0 & \frac{2\sqrt{c}}{1+c} \\ 
0 & -1 & 0 \\ 
\frac{2\sqrt{c}}{1+c} & 0 & \frac{c-1}{1+c} 
\end{pmatrix} ; \quad
\begin{aligned}
&\langle 0|B|0\rangle = -\langle 2|B|2\rangle = \frac{1-c}{1+c} \\
&\langle 2|B|0\rangle = \langle 0|B|2\rangle = \frac{2\sqrt{c}}{1+c} \\
&\langle 1|B|1\rangle = -1
\end{aligned} \]
Then
\[
S_{B_0} = (D_{00} - D_{22}) \frac{1-c}{1+c} - D_{11} + (D_{02} + D_{20}) \frac{2\sqrt{c}}{1+c} \]
\paragraph{CHSH Sum}
\begin{align*}
S_{\rm{CHSH}} &= \langle (R(\omega_2) + R(\omega_0)) \otimes B_{m}B_{m+1} + (R(\omega_2) - R(\omega_0)) \otimes B_0 \rangle \\
&= \langle R(\omega_2) \otimes B_{m}B_{m+1} \rangle + \langle R(\omega_2) \otimes B_0 \rangle + \langle R(\omega_0) \otimes B_{m}B_{m+1} \rangle - \langle R(\omega_0) \otimes B_0 \rangle
\end{align*}
Denote: $S_{m,m+1,\omega_0} - S_{0,\omega_0} =R(\omega_0) \otimes B_{m}B_{m+1} - R(\omega_0) \otimes B_0    $, $S_{m,m+1,\omega_2} + S_{0,\omega_2} = R(\omega_2) \otimes B_{m}B_{m+1} + R(\omega_2) \otimes B_0.$ 
Then, the explicit of $S_{CHSH}$ is:
\[
S_{m,m+1,\omega_0} - S_{0,\omega_0} = (D_{00}^{\omega_0} - D_{22}^{\omega_0}) \left(\frac{-2c}{1+c}\right) + 2D_{11}^{\omega_0} + (D_{02}^{\omega_0} + D_{20}^{\omega_0}) \frac{\sqrt{c}}{1+c} \left( 4(-1)^m s_2 - 2 \right)
\]
\[
S_{m,m+1,\omega_2} + S_{0,\omega_2} = (D_{00}^{\omega_2} - D_{22}^{\omega_2}) \frac{2-4c}{1+c} + (D_{02}^{\omega_2} + D_{20}^{\omega_2}) \frac{\sqrt{c}}{1+c} \left( 4(-1)^m s_2 + 2 \right)
\]
\textbf{First, we calculate $W_0,W_1,W_2$:}
\[\begin{aligned} W_0 &= D_{00}^\omega - D_{22}^\omega = \cos \omega \left( |c_{00}|^2 - |c_{10}|^2 \right) + 2\sin \omega \text{Re}(c_{10}^* c_{00}) \\
& \qquad \qquad \qquad  - \left[ \cos \omega \left( |c_{02}|^2 - |c_{12}|^2 \right) + 2\sin \omega \text{Re}(c_{12}^* c_{02}) \right] \\
\implies W_0 &= \cos \omega \left( |c_{00}|^2 - |c_{10}|^2 - |c_{02}|^2 + |c_{12}|^2 \right) + 2\sin \omega \left( \text{Re}(c_{10}^* c_{00}) - \text{Re}(c_{12}^* c_{02}) \right)
\end{aligned}
\]
And
\[
W_1 = D_{11}^\omega = \cos \omega \left( |c_{01}|^2 - |c_{11}|^2 \right) + 2\sin \omega \text{Re}(c_{11}^* c_{01})
\]
And
\begin{equation}
\begin{aligned}
W_2 &= D_{02}^\omega + D_{20}^\omega = \cos \omega (c_{00}^* c_{02} - c_{10}^* c_{12}) + \sin \omega (c_{10}^* c_{02}+c_{00}^*c_{12}) \\
& \qquad \qquad \qquad   + \cos \omega (c_{02}^* c_{00} - c_{12}^* c_{10}) + \sin \omega (c_{12}^* c_{00}+c_{02}^*c_{10}) \\
\implies W_2 &= \cos \omega \left[ (c_{00}^* c_{02} + c_{02}^* c_{00}) - (c_{10}^* c_{12} + c_{12}^* c_{10}) \right] + \sin \omega (c_{12}^* c_{00}+c_{02}^*c_{10}+c_{10}^* c_{02}+c_{00}^*c_{12})
\end{aligned}
\end{equation}
\textbf{Secondly, we calculate $S_{m,m+1,\omega_0} - S_{0,\omega_0} $:}
\[S_{m,m+1,\omega_0} - S_{0,\omega_0} = W_0^{\omega_0} \left(\frac{-2c}{1+c}\right) + 2W_{1}^{\omega_0} + W_{2}^{\omega_0} \frac{\sqrt{c}}{1+c} \left( 4(-1)^m s_2 - 2 \right)\]
Explicit:
\begin{equation}
    \begin{aligned}
S_{m,m+1,\omega_0} - S_{0,\omega_0} &= \cos \omega_0 \left[ \frac{-2c}{1+c} \left( |c_{00}|^2 - |c_{10}|^2 - |c_{02}|^2 + |c_{12}|^2 \right) + 2\left( |c_{01}|^2 - |c_{11}|^2 \right) \right. \\
&\quad \left. + \frac{2\sqrt{c}}{1+c} \text{Re}(c_{00}^* c_{02} - c_{10}^* c_{12}) \left( 4(-1)^m s_2 - 2 \right) \right] \\
&\quad + \sin \omega_0 \left[ \frac{-4c}{1+c} \text{Re}(c_{10}^* c_{00} - c_{12}^* c_{02}) + 4\text{Re}(c_{11}^* c_{01}) \right. \\
&\quad \left. + \frac{2\sqrt{c}}{1+c} \text{Re}\left( c_{12}^* c_{00} + c_{02}^* c_{10} \right) \left( 4(-1)^m s_2 - 2 \right) \right]
\end{aligned}
\end{equation}
\textbf{Finally, we calculate $S_{m,m+1,\omega_2} + S_{0\omega_2}$:}
\[
S_{m,m+1,\omega_2} + S_{0\omega_2} = W_0^{\omega_2} \frac{2-4c}{1+c} + W_2^{\omega_2} \frac{\sqrt{c}}{1+c} \left( 4(-1)^m s_2 + 2 \right)
\]
\begin{equation}
    \begin{aligned}
S_{m,m+1,\omega_2} + S_{0,\omega_2} &= \cos \omega_2 \left[ \frac{2-4c}{1+c} \left( |c_{00}|^2 - |c_{10}|^2 - |c_{02}|^2 + |c_{12}|^2 \right) \right. \\
&\quad \left. + \frac{2\sqrt{c}}{1+c} \left( \text{Re}(c_{00}^* c_{02} - c_{10}^* c_{12}) \right) \left( 4(-1)^m s_2 + 2 \right) \right] \\
&\quad + \sin \omega_2 \left[ \frac{2-4c}{1+c} \cdot 2 \text{Re}(c_{10}^* c_{00} - c_{12}^* c_{02}) \right. \\
&\quad \left. + \frac{2\sqrt{c}}{1+c} \text{Re}\left( c_{12}^* c_{00} + c_{02}^* c_{10} \right) \left( 4(-1)^m s_2 + 2 \right) \right]
\end{aligned}
\end{equation}
$S_{\rm{CHSH}}$ can be write in form: :
\begin{equation}
\langle S_{\rm{CHSH}}\rangle
=
X_0 \cos\omega_0 + Y_0 \sin\omega_0 + X_2 \cos\omega_2 + Y_2 \sin\omega_2 .
\label{CHSH_appendix}
\end{equation}
$\langle S_{\rm{CHSH}}\rangle|$ max $\iff$ :
\begin{itemize}
    \item $f(\omega_0) = X_0 \cos\omega_0 +
Y_0 \sin\omega_0 $ max $\iff f'(\omega_0) = 0 \iff Y_0 \cos\omega_0 -
X_0 \sin\omega_0 = 0 \iff \omega_0 = \arctan{\frac{Y_0}{X_0}} $
\item Similarly, $w_2 = \arctan{\frac{Y_2}{X_2}}$
\end{itemize}
More explicit: 
\begin{equation}
\begin{aligned}
X_0 &= \frac{-2c}{1+c} \left( |c_{00}|^2 - |c_{10}|^2 - |c_{02}|^2 + |c_{12}|^2 \right) + 2 \left( |c_{01}|^2 - |c_{11}|^2 \right) \\ &\qquad\qquad\qquad\qquad\qquad\qquad\qquad\qquad\qquad  + \frac{2\sqrt{c}}{1+c} \text{Re}(c_{00}^* c_{02} - c_{10}^* c_{12}) \left( 4(-1)^m s_2 - 2 \right) \\
Y_0 &= \frac{-4c}{1+c} \text{Re}(c_{10}^* c_{00} - c_{12}^* c_{02}) + 4\text{Re}(c_{11}^* c_{01}) 
 + \frac{2\sqrt{c}}{1+c} \text{Re}\left( c_{12}^* c_{00} + c_{02}^* c_{10} \right) \left( 4(-1)^m s_2 - 2 \right) 
\end{aligned} \label{X0Y0}
\end{equation}
And
\begin{equation}
\begin{aligned}
X_2 &= \frac{2-4c}{1+c} \left( |c_{00}|^2 - |c_{10}|^2 - |c_{02}|^2 + |c_{12}|^2 \right) + \frac{2\sqrt{c}}{1+c} \text{Re}(c_{00}^* c_{02} - c_{10}^* c_{12}) \left( 4(-1)^m s_2 + 2 \right) \\
Y_2 &= 2 \left( \frac{2-4c}{1+c} \right) \text{Re}(c_{10}^* c_{00} - c_{12}^* c_{02}) + \frac{2\sqrt{c}}{1+c} \text{Re}\left( c_{12}^* c_{00} + c_{02}^* c_{10} \right) \left( 4(-1)^m s_2 + 2 \right)
\end{aligned} \label{X2Y2}
\end{equation}
The optimal value of the CHSH sum is given by:
\begin{equation}
S_{\rm{CHSH}}^{opt} = \max_{\omega_0, \omega_2} \left[ (S_{m,m+1,\omega_0} - S_{0,\omega_0}) + (S_{m,m+1,\omega_2} + S_{0,\omega_2}) \right] = \sqrt{X_0^2 + Y_0^2} + \sqrt{X_2^2 + Y_2^2} \label{CHSH_optimal}    
\end{equation}
\paragraph{KCBS Sum}

To calculate the expectation value of the KCBS operator S , we start with the operator defined as:
\begin{align*}
S &= \sum_{j=0}^{n-2} B_j B_{j+1} - B_{n-1}B_0 = 4 \sum_{j=0}^{n-1} P_j - n I
\end{align*}
In the basis $\{|0\rangle, |1\rangle, |2\rangle\}$ , the operator S is diagonal:
\begin{align*}
S &= \text{diag} \left( n \frac{1-\cos(\pi/n)}{1+\cos(\pi/n)}, n \frac{1-\cos(\pi/n)}{1+\cos(\pi/n)}, n \frac{3\cos(\pi/n)-1}{1+\cos(\pi/n)} \right) \\
  &= \text{diag} (\lambda_1, \lambda_2, \lambda_3)
\end{align*}
Consider state:
$$|\varphi\rangle = \sum_{j=0}^1 \sum_{k=0}^2 c_{jk} |j\rangle|k\rangle$$

 $S = \text{diag}(\lambda_1, \lambda_1, \lambda_3)$, Apply $I \otimes S$ on $|\varphi\rangle$ is:
$$(I \otimes S)|\varphi\rangle = \sum_{j=0}^1 \lambda_1 c_{j0} |j\rangle|0\rangle + \sum_{j=0}^1 \lambda_1 c_{j1} |j\rangle|1\rangle + \sum_{j=0}^1 \lambda_3 c_{j2} |j\rangle|2\rangle$$

The expectation value $\langle I \otimes S \rangle$ can be calculated:
\[
\begin{aligned}
\langle I \otimes S \rangle &= \langle \varphi | I \otimes S | \varphi \rangle \\
&= \lambda_1 \left( \sum_{j=0}^1 |c_{j0}|^2 \right) + \lambda_1 \left( \sum_{j=0}^1 |c_{j1}|^2 \right) + \lambda_3 \left( \sum_{j=0}^1 |c_{j2}|^2 \right) \\
&= \lambda_1 \left( \sum_{j=0}^1 |c_{j0}|^2 + \sum_{j=0}^1 |c_{j1}|^2 \right) + \lambda_3 \left( \sum_{j=0}^1 |c_{j2}|^2 \right)
\end{aligned}
\]
Use normalization condition of probability: $\sum_{j=0}^1 \sum_{k=0}^2 |c_{jk}|^2 = 1$, we receive the final result:
\begin{equation}
\begin{aligned}
S_{\rm{KCBS}}&= \langle I \otimes S \rangle = \lambda_3 \left( \sum_{j=0}^1 |c_{j2}|^2 \right) + \lambda_1 \left( 1 - \sum_{j=0}^1 |c_{j2}|^2 \right) = (\lambda_3 - \lambda_1) \left( \sum_{j=0}^1 |c_{j2}|^2 \right) + \lambda_1 \\
    &= n\frac{4c-2}{1+c}\sum_{j=0}^1 |c_{j2}|^2 + n\frac{1-c}{1+c} \label{KCBS}
\end{aligned}
\end{equation}

\subsection{$B_0$ and $B_mB_{m+1}$}
\[
|\psi_j\rangle =
\frac{1}{\sqrt{1+\cos(\pi/n)}}
\begin{pmatrix}
\cos\!\left(j\frac{(n-1)\pi}{n}\right), &
\sin\!\left(j\frac{(n-1)\pi}{n}\right), &
\sqrt{\cos(\pi/n)}
\end{pmatrix}^T
\]

and

\[
B_j = (-1)^j \left( 2|\psi_j\rangle\langle\psi_j| - I \right).
\]
Let 
\[
c=\cos\!\left(\frac{\pi}{n}\right), \qquad 
c_2=\cos\!\left(\frac{\pi}{2n}\right), \qquad 
s_2=\sin\!\left(\frac{\pi}{2n}\right), 
\]
and let \(m=\frac{n-1}{2}\).

Then the two vectors are
\begin{align*}
|\psi_m\rangle &= \frac{1}{\sqrt{1 + c}} \begin{pmatrix} (-1)^m s_2,&  (-1)^{m+1} c_2,&  \sqrt{c} \end{pmatrix}^T \\
|\psi_{m+1}\rangle &= \frac{1}{\sqrt{1 + c}} \begin{pmatrix} (-1)^m s_2,&  (-1)^m c_2,& \sqrt{c} \end{pmatrix}^T
\end{align*}
First, we calculate: $B_mB_{m+1} = 2(P_m+P_{m+1})-I:$

The projectors \(P_j = |\psi_j\rangle\langle\psi_j|\) give:
\begin{align*}
P_m &= |\psi_m\rangle\langle\psi_m| = \frac{1}{1+c} \begin{pmatrix} s_2^2 & -s_2 c_2 & (-1)^m s_2 \sqrt{c} \\ -s_2 c_2 & c_2^2 & (-1)^{m+1} c_2 \sqrt{c} \\ (-1)^m s_2 \sqrt{c} & (-1)^{m+1} c_2 \sqrt{c} & c \end{pmatrix} \\
P_{m+1} &= |\psi_{m+1}\rangle\langle\psi_{m+1}| = \frac{1}{1+c} \begin{pmatrix} s_2^2 & s_2 c_2 & (-1)^m s_2 \sqrt{c} \\ s_2 c_2 & c_2^2 & (-1)^m c_2 \sqrt{c} \\ (-1)^m s_2 \sqrt{c} & (-1)^m c_2 \sqrt{c} & c \end{pmatrix}
\end{align*}
So:
\begin{align*}
P_m + P_{m+1} &= \frac{1}{1+c} \begin{pmatrix} 2s_2^2 & 0 & 2(-1)^m s_2 \sqrt{c} \\ 0 & 2c_2^2 & 0 \\ 2(-1)^m s_2 \sqrt{c} & 0 & 2c \end{pmatrix}
\end{align*}
Then:
\begin{equation}
    B_mB_{m+1} = \begin{pmatrix} \frac{1-3c}{1+c} & 0 & \frac{4(-1)^m s_2 \sqrt{c}}{1+c} \\ 0 & 1 & 0 \\ \frac{4(-1)^m s_2 \sqrt{c}}{1+c} & 0 & \frac{3c-1}{1+c} \end{pmatrix} \label{BmBm+1}
\end{equation}
Secondely, we calculate $B_0 =2|\psi_0\rangle\langle\psi_0|-I $
For j = 0, $|\psi_0\rangle=\frac{2}{1+c}(1,0,\sqrt{c})^T$, then we have
\begin{equation}
B_0=
\begin{pmatrix}
\dfrac{1-c}{1+c} &0&\dfrac{2\sqrt c}{1+c}\\
0&-1&0\\
\dfrac{2\sqrt c}{1+c}&0&\dfrac{c-1}{1+c}
\end{pmatrix} \label{B0}
\end{equation}

\section{Quantum Circuit} 
In this appendix, we present the explicit circuit constructions underlying the protocol. 
We first explain how the Fourier-test scheme enables direct estimation of expectation values of unitary observables, which forms the basis of our measurement implementation. 
We then provide the detailed circuit realizing the preparation of the minimal nonlocal--contextual state considered in this work.
\subsection{Quantum Fourier Test}
\label{QC_Fourier_Proof} 

As introduced in Eq.~\ref{U_expectation}, we now compute explicitly the outcome of this circuit.

We aim to evaluate a Hermitan and Unitary operator $U$:
\[
\langle U\rangle = \langle \psi | U | \psi \rangle,
\]

The qutrit Fourier transform is defined as
\[
F_3 |k\rangle = \frac{1}{\sqrt{3}} \sum_{j=0}^{2} \omega^{jk} |j\rangle,
\quad \omega = e^{2\pi i / 3}.
\]

Starting from the initial state
\[
|0\rangle |\psi\rangle
\;\xrightarrow{F_3}\;
\frac{1}{\sqrt{3}} \sum_{a=0}^{2} |a\rangle |\psi\rangle,
\]

we apply the controlled operation
\[
|a\rangle |\psi\rangle \;\mapsto\; |a\rangle U^a |\psi\rangle,
\]
where \( U^2 = I \) and \( U^\dagger = U \).
This yields
\[
|\Psi\rangle =
\frac{1}{\sqrt{3}}
\left(
|0\rangle |\psi\rangle
+ |1\rangle U |\psi\rangle
+ |2\rangle |\psi\rangle
\right).
\]

Applying \( F_3^\dagger \), we obtain
\[
|\Psi'\rangle =
\frac{1}{3} \sum_{j=0}^{2}
|j\rangle
\Big(
(1 + \omega^{-2j}) I + \omega^{-j} U
\Big)
|\psi\rangle.
\]
Define
\[
|\phi_j\rangle =
\frac{1}{3}
\Big(
(1 + \omega^{-2j}) I + \omega^{-j} U
\Big)
|\psi\rangle,
\qquad
P(j) = \|\phi_j\|^2.
\]
Using \( U^2 = I \), we obtain
\[
P(0) = \frac{1}{9}(5 + 4\langle U\rangle), \qquad
P(1) = P(2) = \frac{1}{9}(2 - 2\langle U\rangle).
\]
Thus,
\begin{equation}
\langle U\rangle=\frac{9P(0) - 5}{4}=\frac{2 -9P(1)}{2}=\frac{9\big(P(0) - P(1) - P(2)\big) - 1}{8}
\label{U_expectation}
\end{equation}

\subsection{Explicit State Preparation Circuit for the Minimal Nonlocal--Contextual State}\label{Preparingstate1}
We consider a hybrid Hilbert space where Alice’s subsystem is effectively two-dimensional (\(d_A = 2\)), embedded in a qutrit space, while Bob’s subsystem is fully three-dimensional (\(d_B = 3\)). Hence, Alice’s system is restricted to the subspace \(\{|0\rangle, |1\rangle\}\).
The target state is:
\[
|\psi\rangle 
= \sin\frac{\theta}{2}\,|00\rangle 
+ \cos\frac{\theta}{2} e^{i\phi}\,|12\rangle .
\]

Starting from the initial state
\[
|00\rangle,
\]
the preparation proceeds as follows:

\paragraph{(i) Single-qutrit rotation.} 
Apply
\[R_{01}^y(\pi - \theta)
=
\begin{pmatrix}
\sin\frac{\theta}{2} & -\cos\frac{\theta}{2} & 0 \\
\cos\frac{\theta}{2} & \sin\frac{\theta}{2} & 0 \\
0 & 0 & 1
\end{pmatrix}\]
which yields
\[
\sin\frac{\theta}{2}\,|00\rangle 
+ \cos\frac{\theta}{2}\,|10\rangle .
\]

\paragraph{(ii) Phase gate.}
Apply
\[
D(\phi,0) = \mathrm{diag}(1, e^{i\phi}, 1),
\]
resulting in
\[
\sin\frac{\theta}{2}\,|00\rangle 
+ e^{i\phi}\cos\frac{\theta}{2}\,|10\rangle .
\]

\paragraph{(iii) Controlled operation.}
Apply a controlled-\(X_{02}\) gate with
\[
X_{02} =
\begin{pmatrix}
0 & 0 & 1 \\
0 & 1 & 0 \\
1 & 0 & 0
\end{pmatrix},
\]
which prepares
\[
|\psi\rangle 
= \sin\frac{\theta}{2}\,|00\rangle 
+ e^{i\phi}\cos\frac{\theta}{2}\,|12\rangle .
\]

\section{Some Quantum Gates in Qutrit Systems} \label{Gates}
The Lie algebra $\mathfrak{su}(3)$ is generated by the eight Gell-Mann matrices $\{\lambda_a\}_{a=1}^8$. Any single-qutrit unitary can be expressed as \cite{lindon2023complete} and \cite{iliat2025physically}
\begin{equation*}
U = \exp\left(-i \sum_{a=1}^{8} \theta_a \lambda_a \right).
\end{equation*}
To construct physically meaningful gates, we define SU(2)-like rotations acting on two-level subspaces $(i,j) \in \{(0,1),(0,2),(1,2)\}$.

\subsection*{Generators in each subspace}
\begin{align*}
\text{(0,1):} \quad & \lambda_x^{(01)} = \lambda_1,\quad
\lambda_y^{(01)} = \lambda_2,\quad
\lambda_z^{(01)} = \lambda_3, \\[4pt]
\text{(0,2):} \quad & \lambda_x^{(02)} = \lambda_4,\quad
\lambda_y^{(02)} = \lambda_5,\quad
\lambda_z^{(02)} = \tfrac{1}{2}(\lambda_3 + \sqrt{3}\lambda_8), \\[4pt]
\text{(1,2):} \quad & \lambda_x^{(12)} = \lambda_6,\quad
\lambda_y^{(12)} = \lambda_7,\quad
\lambda_z^{(12)} = \tfrac{1}{2}(-\lambda_3 + \sqrt{3}\lambda_8).
\end{align*}
where: \begin{align*}
\lambda_1 &= \begin{pmatrix} 0 & 1 & 0 \\ 1 & 0 & 0 \\ 0 & 0 & 0 \end{pmatrix} & 
\lambda_2 &= \begin{pmatrix} 0 & -i & 0 \\ i & 0 & 0 \\ 0 & 0 & 0 \end{pmatrix} & 
\lambda_3 &= \begin{pmatrix} 1 & 0 & 0 \\ 0 & -1 & 0 \\ 0 & 0 & 0 \end{pmatrix} \\
\lambda_4 &= \begin{pmatrix} 0 & 0 & 1 \\ 0 & 0 & 0 \\ 1 & 0 & 0 \end{pmatrix} & 
\lambda_5 &= \begin{pmatrix} 0 & 0 & -i \\ 0 & 0 & 0 \\ i & 0 & 0 \end{pmatrix} & 
\lambda_6 &= \begin{pmatrix} 0 & 0 & 0 \\ 0 & 0 & 1 \\ 0 & 1 & 0 \end{pmatrix} \\
\lambda_7 &= \begin{pmatrix} 0 & 0 & 0 \\ 0 & 0 & -i \\ 0 & i & 0 \end{pmatrix} & 
\lambda_8 &= \frac{1}{\sqrt{3}} \begin{pmatrix} 1 & 0 & 0 \\ 0 & 1 & 0 \\ 0 & 0 & -2 \end{pmatrix}
\end{align*}
These eight matrices form a basis of $\mathfrak{su}(3)$.
\subsection*{Rotation gates}
We define qutrit rotations as
\begin{equation}
\begin{aligned}
R_{ij}^{x}(\theta) &= \exp\left(-i \frac{\theta}{2} \lambda_x^{(ij)} \right), \\
R_{ij}^{y}(\theta) &= \exp\left(-i \frac{\theta}{2} \lambda_y^{(ij)} \right), \\
R_{ij}^{z}(\theta) &= \exp\left(-i \frac{\theta}{2} \lambda_z^{(ij)} \right).
\end{aligned}
\end{equation}
\subsection*{Some qutrit gates we use in this paper}
\begin{center}
\begin{tabular}{c|c|c}
\hline
Type & Definition & Matrix form \\
\hline
$R_{01}^{y}(\theta)$
& $e^{-i \frac{\theta}{2} \lambda_2}$
& $\begin{pmatrix}
\cos\frac{\theta}{2} & -\sin\frac{\theta}{2} & 0\\
\sin\frac{\theta}{2} & \cos\frac{\theta}{2} & 0\\
0 & 0 & 1
\end{pmatrix}$ \\[15pt]
$R_{02}^{y}(\theta)$
& $e^{-i \frac{\theta}{2} \lambda_5}$
& $\begin{pmatrix}
\cos\frac{\theta}{2} & 0 & -\sin\frac{\theta}{2}\\
0 & 1 & 0\\
\sin\frac{\theta}{2} & 0 & \cos\frac{\theta}{2}
\end{pmatrix}$ \\[15pt]
$R_{12}^{y}(\theta)$
& $e^{-i \frac{\theta}{2} \lambda_7}$
& $\begin{pmatrix}
1 & 0 & 0\\
0 & \cos\frac{\theta}{2} & -\sin\frac{\theta}{2}\\
0 & \sin\frac{\theta}{2} & \cos\frac{\theta}{2}
\end{pmatrix}$ \\[15pt]
Phase gate
& $D(\alpha,\beta) = \mathrm{diag}(1,e^{i\alpha},e^{i\beta})]$
& $\mathrm{diag}(1,e^{i\alpha},e^{i\beta})$ \\
Controlled-Gate & $|a\rangle |\psi\rangle \;\mapsto\; |a\rangle U^a |\psi\rangle,
$ & diag($I_3,U,U^2$) \\
\hline
\end{tabular}
\end{center}
\medskip
These gates realize SU(2) rotations embedded within SU(3), each selectively addressing a two-level subspace of the qutrit. The phase gate $D(\alpha,\beta)$ generates relative phase shifts between basis states. The controlled operation $C^{(a)}(U)$ conditionally applies $U$ to the target qutrit when the control is in state $|a\rangle$.

\section{Detailed analysis of the optimal state in the large-$n$ regime}
\label{large_n_state_I}

In the large-$n$ regime, contextuality and nonlocality coexist within a constrained asymptotic window. As $X_0, X_2 \to -1$ and $Y_0 = -Y_2 \to -\sin\theta\cos\phi$, the CHSH value approaches
\begin{equation}
S_{\mathrm{CHSH}} \to 2\sqrt{1+\sin^2\theta\cos^2\phi},
\end{equation}
which exceeds the classical bound whenever $\sin\theta\cos\phi \neq 0$. Thus, any state with nonvanishing coherence exhibits CHSH violation.

From Eq.~\eqref{thresholdKCBS}, KCBS violation requires $p_2 \to 1 - \frac{2}{n}$. For the state
\[
|\psi\rangle = \sin\frac{\theta}{2}|00\rangle + \cos\frac{\theta}{2} e^{i\phi} |12\rangle,
\]
a small-$\theta$ expansion gives
\[
p_2 = \cos^2\left(\frac{\theta}{2}\right) \approx 1 - \frac{\theta^2}{4}.
\]
Matching this condition yields $\theta \sim \frac{2\sqrt{2}}{\sqrt{n}}$. Therefore, joint violation occurs in the window
\[
\theta \in \left(0, \frac{2\sqrt{2}}{\sqrt{n}}\right), 
\qquad \phi \neq \frac{\pi}{2}.
\]

To identify the optimal coexistence point, we expand both violations near $\theta \approx 0$ and $\phi = 0$ (or $k\pi$). Using Eq.~\eqref{Sum_for_state_I}, we obtain
\[
S_{\mathrm{KCBS}} - (n-2) \approx 2 - \frac{n\theta^2}{4},
\qquad
S^{\mathrm{opt}}_{\mathrm{CHSH}} - 2 \approx \theta^2.
\]
Balancing these two contributions yields
\begin{equation}
\theta^2 = \frac{8}{n+4},
\end{equation}
and hence
\begin{equation}
S_{\mathrm{quantum}} - S_{\mathrm{classical}} \approx \frac{8}{n+4} \sim \frac{1}{n} > 0.
\end{equation}
\subsection{Explicit forms for Eq\eqref{Theorem1_eq}}
\label{explicitTheorem1}
For the state
\begin{equation}
    |\psi_n\rangle = \sqrt{\frac{2}{n+4}}\,|00\rangle + \sqrt{\frac{n+2}{n+4}}\, e^{ik\pi}|12\rangle , \qquad k \in \mathbb{Z},    
\end{equation}
which gives $\sin\frac{\theta}{2} = \sqrt{\frac{2}{n+4}}$ and $\cos\frac{\theta}{2} = \sqrt{\frac{n+2}{n+4}}$, yielding $\sin\theta = 2\sqrt{\frac{2(n+2)}{(n+4)^2}}$.

In the asymptotic limit $n\to \infty$, we assume $c=1$. The KCBS term becomes: 
\begin{equation}
    \begin{aligned}
        S_{\rm{KCBS}} - (n-2) &= \frac{n}{1+c}\left[(4c-2)\cos^2\left(\frac{\theta}{2}\right) - 2c\right] + 2 \\
        &\xrightarrow{n \to \infty} \frac{n}{2}\left(2\,\frac{n+2}{n+4} - 2\right) + 2 \\
        &= \frac{8}{n+4}.
    \end{aligned}
\end{equation}

Similarly, for the CHSH term, by taking $c=1$, $s_2=0$, and $\cos^2\phi=1$, the expression simplifies as follows: 
\begin{equation}
    \begin{aligned}
        S_{\rm{CHSH}}^{\mathrm{opt}} - 2 &= \frac{1}{1+c} \Bigg[ \sqrt{4c^2 + c\sin^2\theta\cos^2\phi\left(4(-1)^m s_2 - 2\right)^2} \\
        &\hspace{4em} + \sqrt{(2-4c)^2 + c\sin^2\theta\cos^2\phi\left(4(-1)^m s_2 + 2\right)^2} \Bigg] - 2 \\[0.5em]
        &\to \frac{1}{2} \left[ \sqrt{4 + \frac{8(n+2)}{(n+4)^2} \cdot 4 } + \sqrt{4 + \frac{8(n+2)}{(n+4)^2} \cdot 4 } \right] - 2 \\[0.5em]
        &= \sqrt{4 + 4\left(\frac{8(n+2)}{(n+4)^2}\right)} - 2 \\[0.5em]
        &\approx 2 \left[ 1 + \frac{8(n+2)}{2(n+4)^2} \right] - 2 \\[0.5em]
        &= \frac{8(n+2)}{(n+4)^2}.
    \end{aligned}
\end{equation}
\end{document}